\newif\ifacm

\ifacm
  \documentclass{src/acm_proc_article-sp} 
\else
  \documentclass{src/llncs}
\fi

\usepackage{color}
\usepackage{xspace}
\usepackage{paralist}
\usepackage{listings}
\usepackage{epsfig}
\usepackage{amssymb}
\usepackage{amsmath}
\usepackage{amsfonts}
\usepackage{wrapfig}
\usepackage{hyperref}

\def \mathtxt #1{$\smash{\mathit{#1}}$}

\newcommand\myparagraph[1]{\medskip\noindent\textbf{#1.}}

\def \fig #1#2#3#4{\begin{figure}[tp]\centering\epsfig{trim=#2, clip=true, width=#3\textwidth, file=res/#1.eps}\vspace{-3mm}\caption{#4\vspace{-3mm}}\label{#1}\end{figure}}

\lstnewenvironment{lsthere}{\lstset{aboveskip=\smallskipamount, belowskip=\smallskipamount, frame=none, columns=fixed, tabsize=2, escapeinside={<latex>}{</latex>}, basicstyle=\scriptsize\ttfamily}}{\vspace{-\parskip}}
\lstnewenvironment{lstfloat}[2]{
  \lstset{label=#1, caption={#2}, float=t, frame=single, captionpos=b, columns=fixed, tabsize=2, belowskip=0px, belowcaptionskip=-10px, escapeinside={<latex>}{</latex>}, basicstyle=\scriptsize\ttfamily, xleftmargin=\dimexpr\fboxsep+\fboxrule+0px, xrightmargin=\dimexpr\fboxsep+\fboxrule+0px}
  \setcounter{lstlisting}{\value{figure}}
  \addtocounter{figure}{1}
  
}{}

\newcommand\cC{\ensuremath{\mathcal{C}}\xspace}
\newcommand\cD{\ensuremath{\mathcal{D}}\xspace}

\newcommand\cP{\ensuremath{\mathcal{P}}\xspace}

\newcommand\cT{\ensuremath{\mathcal{T}}\xspace}
\newcommand\cR{\ensuremath{\mathcal{R}}\xspace}
\newcommand\cU{\ensuremath{\mathcal{U}}\xspace}

\newcommand\SAFE{SAFE\xspace}

\newcounter{examplecounter}

\definecolor{darkgreen}{rgb}{0.1,0.5,0.1}
\definecolor{darkblue}{rgb}{0,0,.6}
\definecolor{shadecolor}{rgb}{0.8,0.8,0.8}
\definecolor{col:form}{rgb}{0.0,0.55,0.73}
\definecolor{col:checkFctn}{rgb}{1,0,0.04}
\definecolor{col:query}{rgb}{0.9,0.18,0.8}
\definecolor{col:credentials}{rgb}{1,0.43,0.13}

\title{Balancing Isolation and Sharing of Data \\ in Third-Party Extensible App Ecosystems}

\ifacm\numberofauthors{3}\fi
\author{
  \ifacm\alignauthor\fi
    Florian Schr\"oder
    \ifacm
      \\
      \affaddr{Saarland University}\\
      \affaddr{Germany}
    \else
      \inst{1}
    \fi
  \and
  \ifacm\alignauthor\fi
    Raphael M. Reischuk
    \ifacm
      \\
      \affaddr{ETH Z\"urich}\\
      \affaddr{Switzerland}
    \else
      \inst{2}
    \fi
  \and
  \ifacm\alignauthor\fi
    Johannes Gehrke
    \ifacm
      \\
      \affaddr{Cornell University}\\
      \affaddr{Ithaca, NY}
    \else
      \inst{3}
    \fi
}

\institute{Saarland University, Germany \and ETH Zurich, Switzerland \and Cornell University, USA}

\begin{document}

\setlength{\abovedisplayskip}{1pt}
\setlength{\belowdisplayskip}{1pt}
\setlength{\abovedisplayshortskip}{1pt}
\setlength{\belowdisplayshortskip}{1pt}

  \maketitle

  \begin{abstract}

In the landscape of application ecosystems, today's cloud
users wish to personalize not only their browsers with
various extensions or their smartphones with various
applications, but also the various extensions and
applications themselves.
The resulting personalization significantly raises the attractiveness for typical Web 2.0 users, but gives rise to various security risks and privacy concerns, such as unforeseen access to certain critical components, undesired information flow of personal information to untrusted applications, or emerging attack surfaces
that were not possible before a personalization has taken place.

In this paper, we propose a novel extensibility mechanism to implement
personalization of existing cloud applications towards (possibly untrusted)
components in a secure and privacy-friendly manner. Our model provides a clean
component abstraction, thereby in particular ruling out undesired component
accesses and ensuring that no undesired information flow takes place between
application components --- either trusted from the base application or
untrusted from various extensions. We then instantiate our model in the SAFE
web application framework (WWW 2012), resulting in a novel methodology that is
inspired by traditional access control and specifically designed for the newly
emerging needs of extensibility in application ecosystems. We illustrate the
convenient usage of our techniques by showing how to securely extend an
existing social network application.

\end{abstract}


\section{Introduction}

In times of massive and still increasing use of web
resources, plat\-form-in\-de\-pen\-dent \emph{Rich Internet
Applications} (RIAs) are considered of high importance.
Called \emph{Software as a Service} (SaaS), these kinds of
applications are often database-driven and predominantly
make high demands on their underlying technology. Today's
Web 2.0 users wish to personalize their devices and
applications -- from minorly invasive \emph{customizations}
(such as changing the visual appearance) to
functionality-extending changes that constitute true forms
of \emph{extensibility}. Not only smartphones, tablets, and
browsers are in focus of personalization, but also existing
RIAs should be customizable -- and even extensible -- in
previously unforeseen
directions~\cite{Arellano12:Personalization,Reischuk12:SAFE,Fredrikson2011:RePriv,Tesoriero2010:CAUCE,Jansen10:CustomizationMultiTenant,Kapitsaki2009:ContextAwareWebApps,Macias2008:CustomizationWebApps,Jorstad06:Personalisation}.

Such user-driven personalizations (sometimes interchangeably
referred to as customizations) inhabit extensible app
ecosystems for web components and influence the
\emph{content}, the \emph{style}, and the
\emph{functionality} of interactive web systems: the welcome
page of Amazon.com shows different items for Alice as
compared to Bob (content), an aged user might wish to have a
larger font size for displaying text on his tablet or
desktop computer (style), while a teenage user might long
for advanced features to publish media data from any
smartphone application to Facebook without waiting for her
OS provider to support the desired features (functionality).
Customization of content and style was traditionally
referred to as personalization in the
literature~\cite{Hardt:12personalization,Toch12:PersonalizationPrivacy,Jorstad06:Personalisation}.
However, with the advent of Web~2.0, \emph{extensibility} of
functionality has become a novel and the most challenging
component in the area of personalization.

One of the central difficulties of realizing extensibility
is to faithfully address the various security and privacy
aspects that naturally arise when functionality is extended
in a user-driven manner. While customization of content and
style usually imposes no security vulnerabilities,
extensibility of functionality (i.e., the
incorporation of new program components into an existing
environment) faces -- apart from the following functional
issues -- also a number of security-related challenges.

(1)~\emph{Functional contracts} between the existing and the
new components have to be met. Consider for example an
address book component $C_A$ that exposes phone numbers to
communication components (e.g., Skype). A specified
personalization could require the address book component
$C_A$ to interact with a \emph{particular} communication component
$C_C$ that might be introduced to the systems by virtue of
extensibility. Functional contracts ensure that the data
exchange format of both interfaces of $C_A$ and $C_C$ match,
i.e., $C_C$ needs to determine not only which global data
exists in the environment of the address book, but also in
which format the data is accessible. Moreover, $C_C$ should
have the option to integrate its own data structures into
the app ecosystem.

(2)~\emph{Security guarantees} have to be ensured for the
entire composed system: (a)~Information flow / privacy:
users want to have credible guarantees that their personal
data is properly protected, they should not be divulged to
potentially untrusted applications or untrusted extensions
of existing applications that were previously considered
trustworthy. The access control policies for the data of the
existing address book component $C_A$ should correctly and
securely be specified when accessed by the additionally
integrated communication component $C_C$. Other components
should securely access data that has been imported by $C_C$
due to the extensibility. (b)~Integrity: users wish to rely
on the integrity of information they get provided, i.e., no
malicious user should be able to interfere in the
communication in a way that alters the result in an
unforeseen or potentially harmful manner. (c)~New attack
vectors: the goal is to augment extensibility with general
security mechanisms that prevent situations in which the
extensibility opens new attack surfaces.
Security is even harder to achieve when new components are
integrated from untrusted and thus potentially malicious
sources. Although software bugs might lead to security holes
in a larger composed system, the chances for an attacker to
introduce malicious components are much higher in open and
extensible environments.

Existing customization frameworks, such as
\cite{Hagemann:2010Mashups,Jansen10:CustomizationMultiTenant,Tesoriero2010:CAUCE,KOL09:WYSIWYG,Kapitsaki2009:ContextAwareWebApps,Ceri2007:ContextAwareWebApps,Bolin:2005CustRendWebPages,Rossi01:PersonalizedWebApps,Danculovic:2001PersWebApps,WebML2000a},
are not suited for our purposes: first, they do not target
security, but solely concentrate on providing proper
functionality; second, they strive for customization rather
than for true extensibility.
We need abstractions for app ecosystems in which the users
can create, share, and install third-party apps through an
``app store'', thereby creating new applications with
enforced security properties.


\myparagraph{Contributions} In a nutshell, this paper
provides a novel mechanism for secure extensibility in the
field of secure web application development. More
precisely, this paper makes the following contributions.

(C1) \textbf{Isolation/Separation.} In order to address the
aforementioned security challenges, we propose a novel
abstraction for controlling access to principal data with a
clear separation for \emph{multiple principal dimensions}.
Our model is inspired by traditional access control models;
however, given the nature of Web 2.0 with extensibility
demands, our model additionally captures the features of
multi-dimensional granularity to support arbitrary
context-aware personalizations and functional extensions. We
provide an instantiation of our model that establishes
enforced data separation in two dimensions: for users and
components of an extensible app ecosystem. This
two-dimensional instantiation provides automatic annotation
of data items to pave the way towards flexible runtime
delegation of privileges and accountability management.


(C2) \textbf{Sharing/Wiring.} Furthermore, in order to share
data across com\-po\-nent/user boundaries, we propose a
wiring methodology to establish explicit data flows between
separated app ecosystem components with explicit control
over the actual data flow. To this end, we have revised the
hierarchical activation model of a recently proposed web
application framework \cite{Reischuk12:SAFE} by a more
sophisticated explicit information flow model for app
ecosystems. Although the activation model nicely corresponds
to the hierarchical structure of HTML web pages,
personalization in terms of true extensibility requires to
move on to a model that allows for data flows beyond the
information propagation along the edges of the hierarchical
data structures proposed in \cite{Reischuk12:SAFE}.

(C3) \textbf{Implementation.} We have implemented our new
extensibility mechanisms in \cite{Reischuk12:SAFE}, which is
a suitable choice for our meth\-od\-ol\-o\-gies since the
framework originally laid the foundation for subsequent
extensions towards secure extensibility. However, the
existing extensibility mechanisms had some architectural
drawbacks, e.g., all application data was globally managed
by a centralized and trusted entity that enforces access
control policies over the data. This notion of global data,
however, does not fully capture the flavor of our
extensibility model and is hard to handle from a security
perspective. The deficiencies have been addressed to provide
more flexibility and enforcement mechanisms to incorporate
the security properties of our model.

(C4) \textbf{Showcase.} Finally, in order to illustrate the
convenient usage of our techniques, we demonstrate how to securely
extend an existing social network web application by an incremental
search functionality that nicely integrates into the previously
existing environment.

\myparagraph{Outline} \autoref{background} provides
background information on traditional access control
mechanisms, and recaps the basics of the underlying
framework. \autoref{sec:model} introduces the abstraction
model for our new mechanisms (C1,\,C2).
\autoref{implementation} presents an implementation of our
mechanisms (C3). \autoref{sec:evaluation} shows the efficacy
of our extensibility approach (C4). \autoref{sec:related}
mentions additional related work and
\autoref{sec:conclusions} concludes.

  \section{Background}
\label{background}

This section provides background information about
traditional access control mechanisms and recaps the basics
of the \SAFE activation framework~\cite{Reischuk12:SAFE}.

\subsection{Access Control Mechanisms}

Traditional access control mechanisms consider the \emph{user} who
requests an operation to decide whether to accept or to reject the
operation on a given dataset. A trusted entity keeps track of
\emph{ownerships} on the datasets that allow for enforcing
appropriate boundaries.
For example, a trusted entity can be the filesystem on a
multi-user desktop computer, which prevents unintended
cross-user file access. \autoref{traditionalAC}a shows a
scenario in which Alice cannot access Bob's home directory,
and vice versa.

Likewise, boundaries across \emph{applications} are
enforced through sandboxes that prevent a particular
application from accessing data in the scope of another
application in a common environment.
As example, a common multi-application scenario that is
suited for deploying such a sandbox is the encapsulation
of application-specific data on contemporary smartphones~\cite{Android}:
In \autoref{traditionalAC}b, the camera software of a
smartphone (component $C_1$) shall not access data stored
by the address book (component $C_2$).

Both concepts have to be combined to achieve extensible and
data-driven access control mechanisms for the recent trends
in cloud applications~\cite{Wang2009:Convergence}, in which
multiple users interact with so-called \emph{mashup}
applications~\cite{Hagemann:2010Mashups} composed of
multiple disjoint software components: It is insufficient to
solely implement per-user access control; data access has to
be additionally restricted to particular software
components. Since third-party software must generally be
considered untrusted, both boundaries have to be enforced
centrally and simultaneously --- we cannot assume any
component to properly and consistently implement user-based
access control for itself.
\autoref{traditionalAC}c depicts the separation of data into two
realms, namely per-user and per-component. Consequently, the central
access control mechanism for any data entity $e$ has to consider at
least the tuple \mathtxt{(uid, cid)}, which can be regarded as the
fixation of two different \emph{access control dimensions}.
\fig{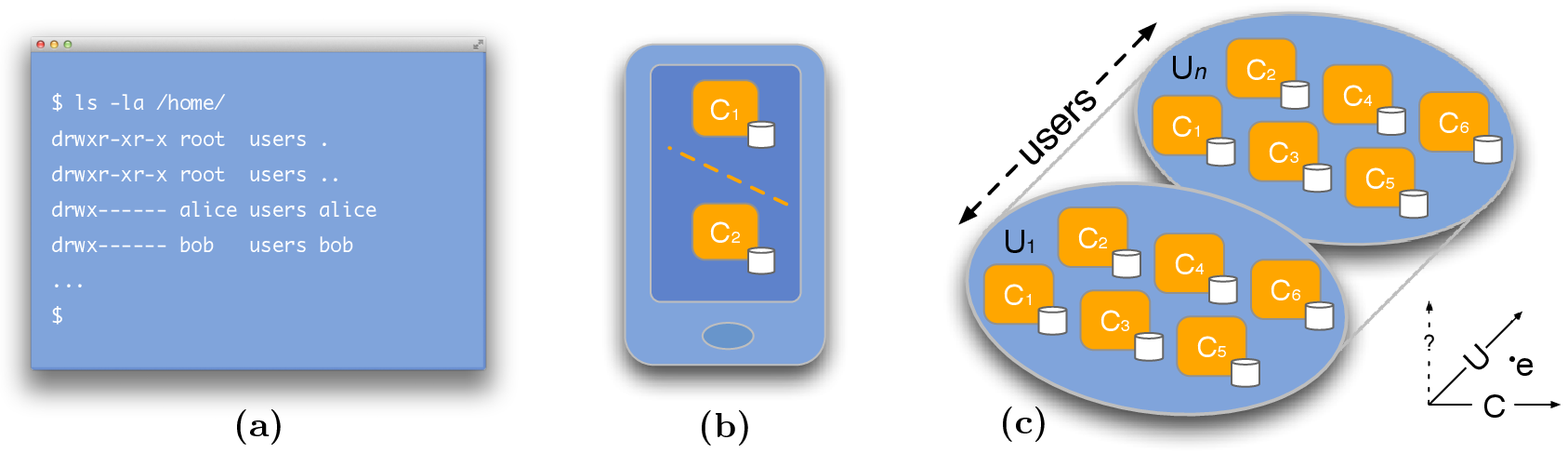}{0px 0px 0px 0px}{0.98}{Data separation (a) on a
multi-user system, (b) on a multi-application system, and (c) in
combination.}

Beyond access control on the basis of two distinct
dimensions, one needs to consider extensibility, modularity,
and personalization, which are crucial properties of modern
RIAs. Usually, these properties imply the possibility of
users who may want to share their own data and particular
applications that may have to jointly operate on the same
datasets, e.g., imagine Alice wants to share her music files with
Bob; likewise, GPS data may be used in both the camera
software component and the address book component. The
resulting need for well-defined interaction amongst users
and/or software components suggests the possibility of
weakening the data separation requirements in either one
dimension, thereby increasing the degrees of freedom beyond
a fixed user or component.

Existing access control approaches are usually
single-dim\-en\-sional. By canonically embedding multiple
dimensions to a single dimension, e.g., by trivial
enumeration of all tuple permutations, one might lose
efficiency and thereby also granularity (see below). A
proper reconciliation of more than one dimension is required
to provide a clean methodology for the design of future
access control policies for app ecosystems.

Moreover, existing approaches with advanced access control
capabilities are not well-suited for modern
web application engineering: Due to the increased
expressivity, approaches such as JIF \cite{Myers1999:JFlow}
or strongly typed languages
\cite{Volpano1996:TypeSystemFlowAnalysis,Fournet:2005:A-Type-Dis,Jia2008:AURA}
usually require explicit annotation, which turns out to be
cumbersome and are thus barely used in heterogeneous
environments formed by independent developers. It is hence
necessary to automatically enforce centrally defined access
control policies rather than relying on user-annotated code.

\subsection{The SAFE Activation Framework}

The \SAFE activation
framework~\cite{Reischuk12:SAFE,Reischuk12:SAFEmanual} is a
web application framework with unified handling of common
techniques largely used by today's RIAs (HTML, CSS, SQL, and
JavaScript). \SAFE is designed for a modularized structuring
of web applications into features, facilitating
extensibility by third-party customizations. The
modularization is achieved by dividing a web application
into semantically coherent features that are provided by
functional self-contained pieces of code, the so-called
\emph{f-units}.



For performance reasons, a web application modeled in \SAFE
maintains a persistent database connection with essentially
non-restricted permissions. The application itself (or its
developer team) is hence responsible for maintaining
well-suited privileges of its users and for enforcing
appropriate security policies such as access control,
information flow, privacy, anonymity, and more.

However, due to \SAFE's open environment with possibly
untrusted f-units, at least the access control policies
across f-units should be enforced by a central and trusted entity in order
to prevent arbitrary data access of potentially malicious
f-units. The trusted entity should provide a generic and
secure interface for defining access policies that reflect
the database semantics as intended by the developers: each
f-unit must be able to rely on the enforcement of the access
control policies that are stated along with the
corresponding f-unit's database tables. Such policies should
be flexible enough to support the ability of extending an
application in unforeseen directions.

  \section{Formal App Ecosystem Model}
\label{sec:model}


This section details a new extensibility concept, presents a
formal model thereof, and provides an instantiation of the
model for the needs of extensible app ecosystems.

\myparagraph{Notation} A common term in the context of
access control is the notion of \emph{principals}.
Principals are usually constraint to users within a system.
Throughout this paper, however, by a principal we denote any
first-class object for which an access control policy may be
applied (e.g., an authenticated user, an installed software
component on a smartphone, or a specific physical location
around a company's headquarters). A principal may possess
and manage its data. A \emph{principal class} is a set of
principals with structurally similar properties (e.g.,
users, software components, devices, locations). We
sometimes refer to the various principal classes as
\emph{principal dimensions}. By $\wp(X)$ we denote the power
set of a set $X$.

\subsection{A Novel Security Extensibility Concept}

The major challenge in defining a suitable principal model
for \emph{extensible} app ecosystems is to develop an
abstraction that satisfies at least the following
requirements.
\textbf{(1)}~The abstraction must take into account the
\textbf{simultaneous interplay of multiple dimensions} (a
user $U$ runs a software component $C$ on a device $D$ at
some physical location $L$, etc.). Note that such an interplay was not
important before the advent of app ecosystems, e.g.,
traditional browser security with extensible plug-ins dealt
with only a \emph{single} user who operates with multi-component
web applications. The security mechanism of an extensible
web application, however, has to take into account \emph{various}
dimensions such as multiple components and multiple users.
\textbf{(2)}~The abstraction must focus on \textbf{efficient
reasoning} for all fields in the cross product of multiple
dimensions. App ecosystems naturally constitute
multi-dimensional \emph{principal grids} in which every principal
class exists in combination with any other principal class.
An ubiquitous access control policy must comprise each cell
in the grid. For each item of the cross product ranging over
all dimensions, a meaningful and efficient policy must
exist. The policy should be concise and transparent since an
embedding of each dimension to single-dimensional
traditional access control policies would not only be
cumbersome to maintain, but might also introduce security
flaws due to the increased complexity of the embedding.
\textbf{(3)}~Extensibility requires the integration of
contextual information in the process of deciding access
control. Dependencies between components and users require
\textbf{context-aware reasoning} methods in which the
context is expressed in terms of a dimension, or by the
presence of information provided by a principal. For
example, owning a certificate might allow a user to access
certain data of a component. Such certificates can be
introduced through extensibility mechanisms and thereby make
the access control mechanisms highly dynamic. Privileges
should not be restricted to (static) binary decisions
(privilege to read data: yes/no), but instead should take
into account an extensible environment with information from
multiple dimensions to allow for more fine-grained and
conditioned policies.

Some of the aforementioned requirements resemble traditional access
control abstractions; others have to be tailored to the specific
needs of extensible web development.
Traditional abstractions for access control (user-based, role-based,
etc.) were tailored to different purposes and are thus constraint to
single dimensions (users, roles, etc.). In the single dimension, only
users are considered first-class citizens; software components are no
first-class objects. Consider, for instance, a UNIX file system in
which Alice's home directory has the permissions \texttt{rwx} (i.e.,
\textbf{r}ead, \textbf{w}rite, and e\textbf{x}ecute) for the owner
Alice (cf.~\autoref{traditionalAC}a). There is no way of specifying
that a particular software component --- in this case some executable
UNIX file --- may access Alice's home directory, while another
component may not. The reason is that components are running on
behalf of users and thus have the same user privileges. However,
components should be treated independently from users, so that
individual access control can be specified in order to deny access to
possibly malicious components (malware, worms, and viruses).
%
%
Moreover, in traditional role-based access control settings, every
component would maintain a list of roles whose users are allowed to
access the component's data. In the file system example, every file
or directory belongs to a group of users. Adding a user to a system
requires to carefully check the user's memberships in the groups of
users. Adding a user to a non-transparent group might grant
unintended privileges to the user.
%

These considerations culminate in a novel abstraction that
is particularly tailored to the emerging paradigm shift in
modern web applications.
The abstraction allows for efficient reasoning and
maintaining the partially conflicting requirements. The
strong forms of extensibility, and in particular the
inter-functionality operation with their mutual conditions
and environmental dependencies, require novel methods that
can be efficiently deployed and maintained.
%
More precisely, in our model, every data item may have an
individual access control policy for every principal in
every dimension.
All principals are thus \emph{first-class citizens} that
inhabit the environment of an extensible web application.
In particular, any principal class can be extended at any
point in time by new principals, e.g., users can be created,
software components can be added, new hardware devices can
be set up, and new physical locations can be explored.
Context-awareness is modeled as part of the extensibility:
the integration of a new component into a system allows for
data integration and the establishment of links to existing
components. This process is referred to as \textbf{wiring}.
A wiring does not only make data flows between components
explicit, but also introduces credentials to state
properties about the actual environment. A credential stated
by component $C_1$ might for instance certify that Alice and
Bob are friends, and hence Bob might read Alice's contact list
which is maintained by a different component $C_2$.
Moreover, our model provides unique ownerships in all
dimensions which can efficiently be inferred by the
currently operating component by the unique position in the
principal grid. As a side product, we believe that this
might help in establishing accountability properties
whenever necessary.
Furthermore, our abstraction contains the concept of
\textbf{sharing} which is based on wirings and ownerships.
The goal of sharing is to provide a reliable mechanism for
enabling explicit information flow across the boundaries of
principals.



\subsection{Multi-Dimensional Principal Model}

We consider the $n$-dimensional universe $\cP^n$ of
\emph{principal classes}
$\cP^n := \langle \cP_1, \ldots, \cP_n \rangle $ that subsumes all
instances of the particular class $\cP_i$, e.g., users,
components, locations, and so on. Furthermore, we define the
\emph{data storage} as the set of all data items \cD. Each
such item is required to have a unique \emph{owner}
principal in each dimension, which would be affected by an
operation on the particular data item. More precisely, for
each data item $d\in\cD$, we define
$
  \mathit{aff_{\cP_i}}\colon \cD \rightarrow \cP_i
$
to represent the \emph{affected principal} in dimension $i$.
The affected principals may be determined with arbitrary
semantics, according to the operation type, information
flow, inference prevention, etc. We intentionally stay as
general as possible here in order to permit a wide range of
possible subsequent instantiations. For instance, items in
\texttt{WHERE} clauses of SQL queries can be captured, or
timing information in the analysis of side-channels.

In order to access data items, a principal can issue a
request $r \in \cR$. We define
$
  \mathit{scope_{\cD}}\colon \cR \rightarrow \wp(\cD)
$
to determine the scope of data items for such a request,
i.e., the set of \emph{affected data items} per request.

As motivated in \autoref{background}, we want to enable
\emph{sharing} between principals of the same dimension,
e.g., user Alice wants to share her favorite music files
with user Bob. We thus require a function
$
  \mathit{sh_{\cP_i}}\colon \cP_i \times \cP_i \times \cD \rightarrow \{0, 1\}
$
for each dimension $\cP_i$
to decide whether sharing from one principal to another is
defined for a specific data item.


Finally, the main access control policy
$
  \mathit{req\_valid}\colon \cR \times \cP_1 \times \ldots \times \cP_n \rightarrow \{0, 1\}
$
decides whether a given request is valid for all principals associated with
this request (the \emph{issuers}).
More specifically, a request $r$ is considered permissive if for
each affected principal $p_i$, we have that either $p_{i}$
itself is the issuer of $r$, or that $p_{i}$ has explicitly shared
the requested data with the actual issuer. Formally,
$\mathit{req\_valid(r, p_1, \ldots, p_n)}$ iff
\begin{align*}
  \begin{aligned}
    \forall d \in \mathit{scope_{\cD}(r)}\colon 
    \bigwedge_{i=1}^{n} ~ \mathit{aff_{\cP_i}(d)} = p_i ~\vee~ \mathit{sh_{\cP_i}(aff_{\cP_i}(d), p_i, d)}.
  \end{aligned}
\end{align*}

\myparagraph{Example}
Consider a set of users \cU and a set of software components
\cC in a web application as an instantiation of two
different principal classes, such that $\cP^2 = \langle \cU,
\cC \rangle$.
If component $c \in \cC$ issues a request $r \in \cR$ on
behalf of user $u \in \cU$, then $r$ is considered
permissive if one of the following conditions holds for all
affected data items $d \in \mathit{scope_{\cD}(r)}$:
\vspace{-\parskip}
\begin{description}
  \setlength\itemsep{-1pt}

  \item[No sharing:] \vspace{-2mm}$\mathit{aff_{\cU}(d)} = u$ and
  $\mathit{aff_{\cC}(d)} = c$, i.e., the request only
  accesses data that is in the scope of both $u$ and~$c$.

  \item[Cross-\cC sharing:] $\mathit{aff_{\cU}(d)} = u$,
  $\mathit{aff_{\cC}(d)} = c'$ for some $c' \not= c$, and
  the component~$c'$ has shared the requested data with
  component~$c$, i.e., $sh_{\cC}(c', c, d)$.

  \item[Cross-\cU sharing:] 
  $\mathit{aff_{\cU}(d)} = u'$ for some $u' \not= u$,
  $\mathit{aff_{\cC}(d)} = c$, and the user~$u'$ has shared
  the requested data with user~$u$, i.e., $sh_{\cU}(u', u,
  d)$.

  \item[Cross-\cU,\cC sharing:] $\mathit{aff_{\cU}(d)} = u'$
  for some $u' \not= u$, $\mathit{aff_{\cC}(d)} = c'$ for
  some $c' \not= c$, and user $u'$ as well as component $c'$
  have both shared the requested data with $c$ running on
  behalf of $u$, i.e., $sh_{\cU}(u', u, d)$ and
  $sh_{\cC}(c', c, d)$.

\end{description}



  \subsection{Instantiation for App Ecosystems}
\label{sec:instantiation}

We show how to instantiate the generic model into a
concrete existing model. This instantiation
constitutes a general role model for extensible app
ecosystems. We will later refine this model to work
within the \SAFE framework.
We concentrate on two dimensions and hence create two
principal classes: authenticated users and software
components.
Finally, we show how to incorporate common relational
database models within our instantiated model. Furthermore,
we show a \emph{wiring methodology} to implement sharing
between components by establishing links between the
database tables owned by the particular components.

Let us reconsider the previous example of a multi-user web
application with extensible components, which
instantiates
$\cP^2 = \langle \cU, \cC \rangle$
as principal universe to constitute the set of users $\cU$ and
software components~$\cC$.

\myparagraph{Tables and Affected Components}
We assume the data storage $\cD$ is reflected by a standard
relational database model. By regarding all data items of
\cD on the granularity of database rows, data items can be
grouped in a set of database tables~\cT expressed as a
relation
$\mathit{data} \colon \cT \rightarrow \wp(\cD).$
%
Database tables are divided into three types (local,
input, output) according to their purpose.
First to mention,
\emph{local tables}
$\mathit{lt} \colon \cC \rightarrow \wp(\cT)$
hold all data items owned by a software component, e.g., all
pictures managed by a photo camera app $\mathit{cam}$ are
stored in its local tables $\mathit{lt}(\mathit{cam})$.
%
Additionally, there is a notion of \emph{input tables}
$ \mathit{it} \colon \cC \times \cU \rightarrow \wp(\cT) $
that subsume data items
which are explicitly provided by other components by means
of \emph{sharing}. For example, the camera app might expect
to find GPS data from the GPS app in an input table
$\mathit{gps}\in\mathit{it}(\mathit{cam},\cdot)$.
Access control policies might impose restrictions on sharing
the actual GPS data according to the user who is accessing
the data. Input tables are therefore instantiated with
respect to a particular user. For example, the camera
application running under user Alice can only access GPS
data for Alice through the input table $\mathit{gps} \in
\mathit{it}(\mathit{cam},\mathit{alice})$.

\myparagraph{Requests}
Database tables (whether local or input) establish a
relation between requests~\cR and concrete data items $\cD$
stored in the respective tables. We define the tables
affected by a request as its scope:
$ \mathit{scope_{\cT}} \colon \cR \rightarrow \wp(\cT) $.
The datasets of tables in the scope include the actually
affected datasets. We hence lift $\mathit{scope_\cT}$ to data
items: we say $d \in \mathit{scope_{\cD}(r)}$ if and only if
$\exists\,t \in \mathit{scope_{\cT}(r)}$ such that $d \in
\mathit{data(t)}$.

We assume that every data item can be accessed by a local
table or by an input table:
$
    \forall d \in \cD \,\; \exists c \in \cC \colon (\exists t \in \mathit{lt(c)} \colon d \in \mathit{data(t)}) \:\vee 
                          (\exists u \in \cU, \exists t \in \mathit{it(c, u)} \colon d \in \mathit{data(t)})
$.

To allow for greatest
flexibility, the content of an input table $t \in
\mathit{it(\cdot, \cdot)}$ may be provided by multiple
different software components at the same time (the details
are presented below). The component providing a particular
data item $d \in \mathit{data(t)}$, i.e., its source
$\mathit{src} \colon \cD \rightarrow \cC$, serves as the
affected component for access on that input table:
$
    \forall t \in \mathit{it(\cdot, \cdot)}, \, d \in \mathit{data(t)} \colon \mathit{aff_{\cC}(d)} := \mathit{src(d)}
$.
For access on local tables, the affected component is the
associated component $c \in \cC$ itself:
$
    \forall t \in \mathit{lt(c)}, d \in \mathit{data(t)} \colon \mathit{aff_{\cC}(d) := c}
$.

\myparagraph{Owners and Affected Users}
Similar to defining the owner of a data item in terms of
software components $\cC$, we next define the owner of a
data item in terms of users $\cU$. In order to determine the
affected user of any access on a data item $d \in \cD$, we
require the presence of an
\emph{owner} mapping
$
  \mathit{own} \colon \cD \rightarrow \cU
$.
Such mapping is expected to be automatically stored with the
data item. The retrieval of the owner information could, for
example, rely on a unique identifier or a particular
owner column for each data item.
By assuming a proper owner information management for both local
tables and input tables, we instantiate the affected user accordingly:
$
    \forall d \in \cD \colon \mathit{aff_{\cU}(d)} := \mathit{own(d)}
$.

\myparagraph{Sharing}
The goal of sharing is to provide a reliable mechanism for enabling
explicit information flow across the boundaries of principals,
thereby enforcing various dynamic confidentiality policies.
In an extensible app ecosystem, we assume that every persistently
stored data item might be processed arbitrarily by a component before
eventually being stored in a particular local table. As components
``see'' any dataset upon insertion, every information represented in
a local table might be reconstructible by the corresponding
component. It is thus irrelevant with respect to confidentiality,
whether we explicitly allow components to directly access arbitrary
data items of an associated local table or not. Moreover, access
control directly on top of local tables could be achieved by
pa\-ram\-e\-ter\-ized views that exclusively provide user-dependent
restricted access to the underlying tables. This so-called
\emph{Truman model}~\cite{Rizvi2004:QueryRewriting} suffers from
various drawbacks. For instance, transparent views that hide
particular data items may introduce subtle inconsistencies for
aggregate functions such as \texttt{AVG} or \texttt{COUNT}.

Due to the limited gain and the anticipated problems of a restriction
directly on local tables, we allow a component~$c$ to have full
access to its local tables for all users. We hence assume for all
local tables $t\!\in\!\mathit{lt(c)}$,
\begin{equation}\label{eq:local_is_public}
    \begin{aligned}
      \forall d\!\in\!\mathit{data(t)}, u\!\in\!\cU \setminus\!\mathit{own(d)}\colon 
     \mathit{sh_{\cU}(own(d), u, d)} 
  \end{aligned}
\end{equation}
Making local tables public in their component's scope does not
introduce potential information leakages 
since components do not gain any additional knowledge. In addition,
the potential leakage or abuse of information has to be considered
anyhow by the user before providing sensitive data to a particular
component.


  Similarly, a user has to rely on the access control mechanisms of
  the providing component in which datasets might be included in an
  input table.
  Thus, given component $c \in \cC$ and user $u \in \cU$, we assume
  for all input tables $t\!\in\!\mathit{it(c, u)}$,
  \begin{equation}\label{eq:input_is_public}
    \begin{aligned}
      \forall d\!\in\!\mathit{data(t)}, u\!\in\!\cU\setminus\!\mathit{own(d)}\colon
                                                                   \mathit{sh_{\cU}(own(d), u, d)} 
    \end{aligned}
\end{equation}
By the definition of an input table $t$, all data items $d \in
\mathit{data(t)}$ are intentionally shared on behalf of the providing
component $\mathit{src(d)}$. Hence, for all input tables of component
$c \in \cC$ running in the scope of user $u \in \cU$, we assume
%
$
    \forall t \in \mathit{it(c, u)}, d \in \mathit{data(t)} \colon \mathit{sh_{\cC}(src(d), c, d)}
$
  %

By Equations (\ref{eq:local_is_public}) and (\ref{eq:input_is_public}), we basically incapacitate the user by shifting the sharing responsibility solely to the component dimension --- in contrast to the requirement that both dimensions have to agree on a sharing of a particular data item.
However, a component can only share datasets that it was explicitly provided with by either the dataset's owner or by another sharing component. We can regard both cases as the implicit affirmation of the user based on his personal trust assessment for potential sharing in a manner the component may specify on own behalf.

\myparagraph{Overall Sandbox}
Using the previously introduced predicates, we propose an
instantiation of the formal model that we call \emph{sandbox}. The
sandbox $\mathit{sb}$ discriminates a particular request according to
common operations in relational database systems:
$
 \mathit{op} \colon \cR \rightarrow \{\mathtt{SEL}, \mathtt{INS}, \mathtt{UPD}, \mathtt{DEL}\}
$.%
\footnote{Here, $\mathit{op}$ is defined to restrict a single request to a
single operation. We stress that this is not necessarily the case in
practice, e.g., an update operation \texttt{UPD} might contain a
select operation \texttt{SEL}.
However, we assume
$\mathit{op(r)}$ to be well-defined for all $r \in \cR$ --- if
necessary, $r$ has to be split up into sub-requests.}
%
We define the semantics of the sandbox
$\mathit{sb} \colon \cR \times \cU \times \cC \rightarrow \{0, 1\}$
as follows: $\mathit{sb(r, u, c)} \mapsto$
\begin{equation}\label{eq:sb}
    \begin{aligned}
    & \;\mathrm{if~}\mathit{op(r)}\!\in\!\{\mathtt{INS}, \mathtt{UPD}, \mathtt{DEL}\}\!\!:
    \underbrace{\forall t\!\in\!\mathit{scope_{\cT}(r)} \colon t\!\in\!\mathit{lt(c)}}_{\text{(\ref{eq:sb}A)}} \wedge
      \underbrace{\forall d\!\in\!\mathit{scope_{\cD}(r)} \colon \mathit{own(d)} \!=\! u}_{\text{(\ref{eq:sb}B)}} \\
    & \;\mathrm{else~if~}\mathit{op(r)} \!\in\! \{\mathtt{SEL}\}\!\!:
    \underbrace{\forall t \in \mathit{scope_{\cT}(r)} \colon t \in \mathit{lt(c)} \vee t \in \mathit{it(c, u)}}_{\text{(\ref{eq:sb}C)}}
    \end{aligned}
\end{equation}
Intuitively, a modification request \{\texttt{INS,UPD,DEL}\} is
considered permissive if it operates only on own local tables
(\ref{eq:sb}A) and if all affected datasets are owned by the
authenticated user (\ref{eq:sb}B). A select request \{\texttt{SEL}\}
is considered permissive, if it operates only on own local tables or
on input tables (\ref{eq:sb}C).

\myparagraph{Soundness}
In order to show that the presented sandbox semantics is a valid
instantiation of the previously introduced formal model, i.e., the
sandbox indeed reflects that for every dimension, either own or
explicitly shared datasets are affected, we have to prove the
soundness of our instantiation with respect to the formal model.
More precisely, it has to be shown that $\mathit{sb(r, u, c)}
\Rightarrow \mathit{req\_valid(r, u, c)}$, i.e., that the sandbox
is at least as restrictive as the model. The implications of the
sandbox semantics immediately entail this statement without further
proof obligations.

\subsection{Component Wiring Model}

One of the major features and challenges of today's data-driven and
reactive web applications 
is to ensure server-client \emph{consistency}. If a component
modifies the state of the database, the changes should be reflected
by all its dependent components (and their visual presentation) and also by the client instances of the components.
In \SAFE, any component (referred to as f-unit) can \emph{activate} other components by means of an activation $ \mathit{act} \colon \cC \rightarrow \wp(\cC)$.
Upon an activation initiated by a component $c$, activation data is passed from $c$ through the particular activation interfaces of $\mathit{act}(c)$. The behavior of the activated child component instances thus depends on the state of the parent component $c$. Consequently, the set $\mathit{act(c)}$ is possibly \emph{data-dependent} of $c$.

F-units generate HTML content that is enclosed by a particular
node in the DOM tree of the HTML page. F-units and their
activations hence constitute a hierarchical, cycle-free structure,
the \emph{activation tree}:
$ 
  \mathit{G_{act}} = \langle \mathit{V_{act}}, 
  \mathit{E_{act}} \rangle = \langle \cC, \{(c, c') \in \cC 
  \times \cC \mid c' \in \mathit{act(c)}\} \rangle
$.
Due to the activation data dependencies, each change in a component's
data realm possibly outdates some components in the corresponding
subtrees of $G_{\mathit{act}}$.

The concept of sharing introduces an additional possibility of
receiving data such that propagation of changes is not necessarily
fully reflected by the edges of the activation tree. This additional
data dependency, as imposed by our sharing mechanism, is covered by
the \emph{combined graph}
$
  \mathit{G_{comb}} = \langle \mathit{V_{act}}, 
  \mathit{E_{act}} \cup \mathit{E_{sh}} \rangle
$
that includes edges representing the presence of an input table from
one component to another:
$  \mathit{E_{sh}} = \{(c, c') \mid \exists u \in \cU, t 
  \in \mathit{it(c', u)}, d \in \mathit{data(t)} \colon c 
  = \mathit{src(d)}\}
$

The definitions of both $\mathit{E_{act}}$ and $\mathit{E_{sh}}$ can
be considered an over-approximation since they do not respect the
extent of the actually changed data --- their combination, however,
clearly captures all possible dependencies.

Using the combined graph, we can determine (and then update)
components that rely on stale data. If a local table of a particular
component $c$ changes due to a modifying query, the transitive
closure starting at $c$ contains all potentially stale components
that should be considered for updating. For the combined graph, we
determine a topological component ordering, which takes the partial
orderings as defined by the particular dependencies into account. The
global topological ordering is well-defined, as wirings and
activations that would result in a cycle are rejected in the first
place, and thereby ensures that the rebuilding step propagates on yet
refreshed data. Using the partial ordering, the order of required
activations (and thus the components to be rebuilt) can be fully
determined --- in other words, all outdated components are rebuilt in
an order such that all data requirements are satisfied. The freshly
generated content is merged into complete subtrees of the activation
tree, and thus also into the DOM tree. Finally, the rebuilt content
is pushed to stale client browser instances.

  \section{Implementation}
\label{implementation}

This section presents the implementation of the major parts of our
model instantiation in the \SAFE framework. More specifically, the
semantics of the sandbox formalisms (\autoref{eq:sb} and its
components 3A, 3B, 3C) rely on the following insights as introduced
below:
\vspace{-\parskip}
\begin{itemize}
  \setlength\itemsep{0pt}

\item[(\ref{eq:sb}A)] \vspace{-1mm} Permissions for \texttt{INS}, \texttt{UPD}, and
\texttt{DEL} operations are only granted on local tables. In
addition, the \emph{query sandbox} (explained below) ensures that a
component can only access its own local tables.

\item[(\ref{eq:sb}B)] According to the \emph{owner invariant}
(explained below), before each modifying operation, MySQL triggers
verify an owner column of the particular row according to the
authenticated user.

\item[(\ref{eq:sb}C)] Permissions for \texttt{SEL}: as for
(\ref{eq:sb}A), the \emph{query sandbox} ensures that only own tables
(whether local or input table) are accessible via SELECT queries.

\end{itemize}
\vspace{-\parskip}
For automatic database management, a \SAFE f-unit may provide
an SQL-style \texttt{.db}-file which declares its tables (local,
input, output). In the following, we present
chosen aspects of the parsing, validation, and interpretation of
\texttt{.db}-files during the f-unit integration process.
We assume the deployment of the widely used open source database
MySQL version 5.1. Furthermore, due to space constraints, we omit
technical details whenever possible.
%
Examples are shown in \autoref{sec:evaluation}.

\myparagraph{Owner Invariant}
We consider a data modification attempt as authorized only if the
operation either adds a new dataset with valid user information or
modifies (or deletes) a dataset that was created on behalf of the
same user before. This achieves separation in the
user dimension (as required by the formal model) as well as the
$\mathit{own(\cdot)}$ validation (as required by part (\ref{eq:sb}B)
in our instantiation). As this approach requires keeping track of the
creating user and the extent of each dataset, we require each dataset
-- more technically, each row of a table -- to hold an \texttt{owner}
column.

In order to ensure accountability, owner-preserving integrity
invariants are defined as transition constraints for each
modification operation on the basis of \texttt{owner} column values:
The \texttt{owner} column of the dataset to be inserted, updated, or
deleted must match the authenticated user the f-unit is currently
connected to. In addition, the \texttt{owner} column must not change
due to an update operation.

For implementing these requirements, MySQL's \emph{trigger} concept
\cite{mysql-create-trigger} is a suitable choice. Before each
particular operation, a trigger inspects a column's pending new and
old value (where appropriate) by the \texttt{NEW} or \texttt{OLD}
pseudo-table, respectively. In addition, direct access to a column
value avoids the need for parsing the query string and thus reduces
the risk for the check of being bypassed.
\autoref{example_trigger.sql} shows an \texttt{UPDATE} trigger that
ensures both requirements of our stated invariant by raising an error
if the value of the \texttt{owner} column has either changed or
cannot be validated against the connected user. The purpose and
semantics of the function \texttt{verify\_uid()}, its arguments, and
the variable \texttt{@uid} are derived in the following. 
The operator \texttt{<=>} is MySQL's NULL-safe equality operator.
\begin{lstfloat}{example_trigger.sql}{MySQL trigger that will be executed before any \texttt{UPDATE} operation on the particular table. Modifications of the \texttt{owner} column are prevented.}
  CREATE TRIGGER tbl_upd_t BEFORE UPDATE ON tbl
    FOR EACH ROW CALL assert(
      NEW.owner<=>OLD.owner AND NEW.owner<=>@uid AND verify_uid('funit','sk')
    );
\end{lstfloat}

To let a trigger verify the owner invariant, the user $u$ known at
\SAFE's centralized reference monitor (CRM) must be made available to
the trigger in a flexible though authentic way. When relying on the
fact that there is a single database connection per CRM and a single
CRM per f-unit processing lifetime, we hence have a single database
connection per f-unit and user. This allows the usage of a
connection-specific MySQL session variable
\cite{mysql-user-variables} to pass the current user $u$ to the
trigger with each query. After establishing the connection to the
database, the CRM thus sets two session variables
$\mathtt{@uid} := \mathit{u}$ and
$\mathtt{@uid\_h} := H(\mathtt{@uid} \mid \mathit{funit} \mid \mathit{sk})$
with the f-unit name $\mathit{funit}$, a secret key $\mathit{sk}$, and a
cryptographic hash function $H$.
  
%

Before the query is executed, the \texttt{verify\_uid()} function in
the trigger of \autoref{example_trigger.sql} is thus able to compare
\texttt{@uid\_h} with the outcome of its own hash computation using
\texttt{@uid}. The included $\mathit{sk}$ inside the hash of
\texttt{@uid\_h} prevents an f-unit from creating valid hashes for
arbitrary users on its own, as the $\mathit{sk}$ is only available to
the CRM and hard-coded in the trigger. Consequently, no f-unit should
be granted the \texttt{TRIGGER} or \texttt{SUPER} privileges
\cite{mysql-show-triggers}. The $\mathit{funit}$ string ensures that
even in case the \texttt{@uid\_h} is leaked, the disclosure is
limited to the scope of the particular f-unit and user.

Each table stated in a \texttt{.db}-file is thus forced to specify
exactly one \texttt{owner} column. This convention allows for the
creation of appropriate triggers that verify this particular column
against the \texttt{@uid} variable that was set by the CRM prior in
the connection --- and thereby enforce the invariants as specified
above.

\myparagraph{Query Sandbox}
Apart from data separation in the user dimension, the formal model
requires a clear data separation between components --- more
technically, the formal model requires an explicit assignment between
tables and f-units, and the prevention of any cross-references. We
thus have to ensure that incoming queries only access tables in the
scope of their originating f-unit.

In order to prevent clashes in the table namespace, every stated
table in an f-unit's \texttt{.db}-file is \emph{prefixed} with the
name of the defining f-unit. For the sake of a convenient usage and a
clear interface, we do not expose the prefixing to the developer.
Instead, the CRM replaces each encountered table in a received query
on-the-fly by its prefixed counterpart. As each f-unit has
authenticated itself at the CRM before placing queries, the f-unit
can be determined reliably. Accessed tables are hence enforced to be
permissive according to the particular connected f-unit. In other
words, the table prefixing thus prevents data access across f-unit
boundaries and thereby implements the \emph{query sandbox}.

In fact, the prefixing approach can be considered as the
transformation of a \emph{global} (shared) database towards a
\emph{local} (per-f-unit) database. The security of the prefixing
approach solely relies on the robustness of the replacing algorithm.

\myparagraph{Wiring}
Due to the limitation of f-units to access only their associated
tables using the query sandbox, we considerably lose flexibility:
cross-f-unit collaboration via the database is prevented, a
contradiction to the extensibility paradigm of \SAFE. We thus have
to provide an implementation of \emph{sharing} using input tables
(cf.~\autoref{sec:instantiation}). We need well-defined interfaces
for exchanging data across f-unit boundaries, while preserving all
integrity and confidentiality constraints.

In order to receive arbitrary data, f-units declare \emph{input
tables} with a table-like signature. \emph{Output tables} implement
\texttt{SELECT} statements for providing such datasets, allowing
f-units to decide on their own, which data shall be exposed.
\autoref{example_iotables.sql} shows an example of an f-unit
providing \emph{user groups} (left-hand side), in which each public
group with its owner is exposed. A \emph{statistics} f-unit
(right-hand side) can receive data items of various types.
As the representation of the data in the providing f-unit does not
necessarily match the intended signature of the input tables of the
receiving f-unit, we should be careful in not limiting the power of
an output table when collecting its information from other tables.
Therefore, arbitrary queries are allowed in the specification of
output tables. However, as with all other f-unit queries, output
table queries are automatically table-pre\-fixed and thus restricted
to the boundaries of the source f-unit. Implemented as a
\texttt{VIEW}, an output table's signature (column names and types)
can be determined reliably after creation using MySQL's
\texttt{information\_schema.columns} table
\cite{mysql-columns-table}. Together with the definitions of input
tables, we can provide full signatures of both input and output
tables to a new step in the integration process, the \emph{wiring}.
\begin{lstfloat}{example_iotables.sql}{Example: Defining input and output tables in an f-unit's \texttt{.db}-file.}
  OUTPUT TABLE all_groups =                INPUT TABLE stats (
    SELECT gid AS key, owner, name           key    KEY
    FROM   groups                            owner  OWNER
    WHERE  public=1                          type   TINYTEXT )
\end{lstfloat}

A \emph{wiring} matches an input table schema to an output table
schema, yielding a particular mapping that is internally expressed as
a \texttt{SELECT} statement. An input table view can thus be
represented by a \texttt{UNION}, which allows to combine multiple
mapping statements --- an approach in data integration terms usually
referred to as \emph{global-as-view}
\cite{Lenzerini2002:DataIntegration}. Each of those input table- and
wiring-specific output table queries form a \emph{schema matching}
that follows syntax and semantics as defined by the input table.
There exist several schema matching techniques that could be used for
automatically deriving input/output table correspondences --- these
techniques are still prone to mistakes, suggesting at least a
human-aided approach \cite{Bernstein2011:SchemaMatching}. However, we
leave further improvements of the wiring process between input and
output tables, such as an algorithm-aided schema matching, for future
work.
Upon integration of an f-unit, the human integrator is presented the
list of all input and output tables and may connect their particular
columns after reviewing types and semantics
(cf.~\autoref{screenshots}, left).

\myparagraph{Foreign Keys}
The goal of input tables is not only to collect data for
presentation, but instead also to extend existing datasets according
to the functionality of an f-unit --- by linking own entries to
received entries. As an example, consider an f-unit managing
particular objects (images, groups, profiles, \ldots), while a wired
child f-unit provides some per-user functionality on top of each
parent object (comments, votes, \ldots).
%
This 1:1 or 1:N dependency can be expressed as values in a local
table that are explicitly referencing a value in another local table
or even in an input table.
Upon deletion of the referenced value, all referencing entries that
have become stale are implicitly deleted in order to ensure
consistency between both involved tables, e.g., if an image has been
deleted, all associated comments are deleted as well.

MySQL's concept of \emph{foreign key} constraints \cite{mysql-innodb-foreign-key-constraints} can be used only for involved tables that are ``real'' tables. However, we want to support foreign keys on input tables, which are implemented as a \texttt{UNION} over the column-mappings to arbitrarily crafted output tables. We thus implement a custom approach in order to emulate foreign key semantics with increased flexibility. Using MySQL triggers, our approach keeps track of dependencies between local tables even through wired input and output tables, and ensures consistency by detecting and deleting rows that have become invalid. Such child rows are not necessarily owned by the user who is deleting the parent object. The operation might hence violate the owner invariant. An explicit handling for such well-defined cases is  implemented as well.

\myparagraph{Wiring Invariants}
By the presence of a wired input table, we assume that the wired
content was shared \emph{intentionally} by the providing source
f-unit. To provide stronger security even in case of such intentional
disclosure, an input table should be restrictable to a particular
user.
As an example, assume an f-unit $F$ that provides the functionality
of friendships between users. While friendship information might be
valuable for other components, e.g., for an f-unit that provides some
messaging feature, each particular dataset of the corresponding
output table of $F$ shall only be accessible for either one of the
involved users. This intuition reflects that a user who provides some
information to $F$ has to trust $F$ in implementing appropriate
access control, whether in the scope of business logic or output
tables.

\begin{lstfloat}{example_otable_ac.sql}{Invariant examples for output tables, using the relation \texttt{is()}, column references, and the variable \texttt{@uid}.}
  OUTPUT TABLE friends_o (
    SELECT    CONCAT(uid1,uid2) AS key,  uid1 AS owner,  uid2 AS friend
    FROM      friends
    INVARIANT is(@uid,owner) OR is(@uid,friend) )
\end{lstfloat}
We thus introduce a deviant output table syntax to incorporate the
possibility of expressing additional \emph{invariants} for each
output table.
As shown in \autoref{example_otable_ac.sql}, the invariant of
\texttt{friends\_o} uses the session variable \texttt{@uid}, the
built-in \emph{predicate} \texttt{is()}, and the logical operator
\texttt{OR}.
The invariant holds true if the \texttt{@uid} matches either the
\texttt{owner} or \texttt{friend} column of the row to be read.
In particular, consider the output table \texttt{friends\_o} being
wired into an input table of a malicious f-unit $F'$. Due to the
invariant of \texttt{friends\_o}, $F'$ would gain knowledge of all
friends of a particular user $u$ only if $u$ has used $F'$ once.
In other words, $F'$ has no access granted to $u$'s datasets until
activated in the scope of $u$'s \texttt{@uid}.
Unless overridden by explicit invariant specification, the default
behavior of output tables assumes the invariant
\texttt{is(@uid,owner)} and thereby protects private data per default.

\myparagraph{Dynamic Predicates} As all information an f-unit might refer to is available by either local tables or by input tables, an f-unit may specify an invariant using tables as predicates. If the particular f-unit considers friendship information retrieval to be permissive for  friends of either one involved party, the invariant could be stated as
\begin{lsthere}
   is(@uid,owner) OR friends(@uid,owner) OR is(@uid,friend) OR friends(@uid,friend)
\end{lsthere}
for \texttt{friends} being a table with binary arity. Likewise, for \texttt{ignores} being an input table or a local table, an invariant consisting of its negation
\begin{lsthere}
   !ignores(owner,@uid) AND !ignores(friend,@uid)
\end{lsthere}
hides datasets for users who are tagged \emph{ignored} by some of the affected users.

For the verification of invariants, the wiring provides an additional
view that selects from the actual (unrestricted) output table.
The \emph{restricted} view of the output table includes a
\texttt{WHERE} condition that is derived from the whole invariant
expression, e.g, from the predicate $\mathtt{tbl(x_0, \ldots, x_i)}$:
\begin{lsthere}
   EXISTS(SELECT * FROM <latex>$\mathit{tbl}$</latex> WHERE <latex>$p_0$</latex>=<latex>$x_0$</latex> AND... AND <latex>$p_i$</latex>=<latex>$x_i$</latex>)
\end{lsthere}
The implementation of invariants containing negation or wildcards is
analogous.


The existential quantifier semantics with conjunctive matching allows
for easy deployment in common environments in which access control is
based on group memberships, permissions, and/or user relationships.
As predicates can even refer to input tables, the wiring process
provides the flexibility and modularity needed for incorporating
extensions at runtime --- knowing the input table's column semantics
is sufficient for an f-unit to state meaningful invariants.

  \section{Examples and Evaluation}
\label{sec:evaluation}

We illustrate how to conveniently extend an existing application with
new functionality, based on the previously introduced techniques.
More specifically, we take a \SAFE application of an interactive
social network and add an incremental search functionality. The search
functionality is modeled as a set of independent f-units.

\myparagraph{Initial Application}
In addition to various other features, the initial social network
application comprises the common functionality of \emph{group
membership} which is implemented by an f-unit \emph{Groups}. Any
authenticated user may create a group, which can be joined by other
users.
As f-units are required to state appropriate output tables for the
sake of extensibility, \emph{Groups} provides the public output table
\texttt{all\_groups} with a declaration of data and one invariant:
\begin{lsthere}
    OUTPUT TABLE all_groups (
      SELECT name, gid AS key, owner FROM groups
      INVARIANT ALL )
\end{lsthere}
The output table exposes the group names to the wiring process: If
wired to \texttt{all\_groups}, other f-units can access the names of
all available groups. The invariant \texttt{ALL} makes the group
information public, i.e., readable for every user, and thus for every
\texttt{@uid}.

Furthermore, an f-unit \emph{Messaging} implements an instant
messaging functionality and defines an output table
\texttt{private\_msgs} as the set of all messages (local table
\texttt{conversations}) that can be associated with the current user:
\begin{lsthere}
    OUTPUT TABLE private_msgs (
      SELECT    msg_id AS key,  msg,  uid_from AS owner,  uid_recipient AS to
      FROM      conversations
      INVARIANT is(owner,@uid) OR is(to,@uid) )
\end{lsthere}
Per default, every user may access output table rows with a matching
owner column \texttt{is(owner,@uid)}, see \autoref{implementation}.
However, the specified invariant replaces this default behavior by
potentially allowing foreign f-units to access both sent and received
messages of the particular user they are currently connected to.


\fig{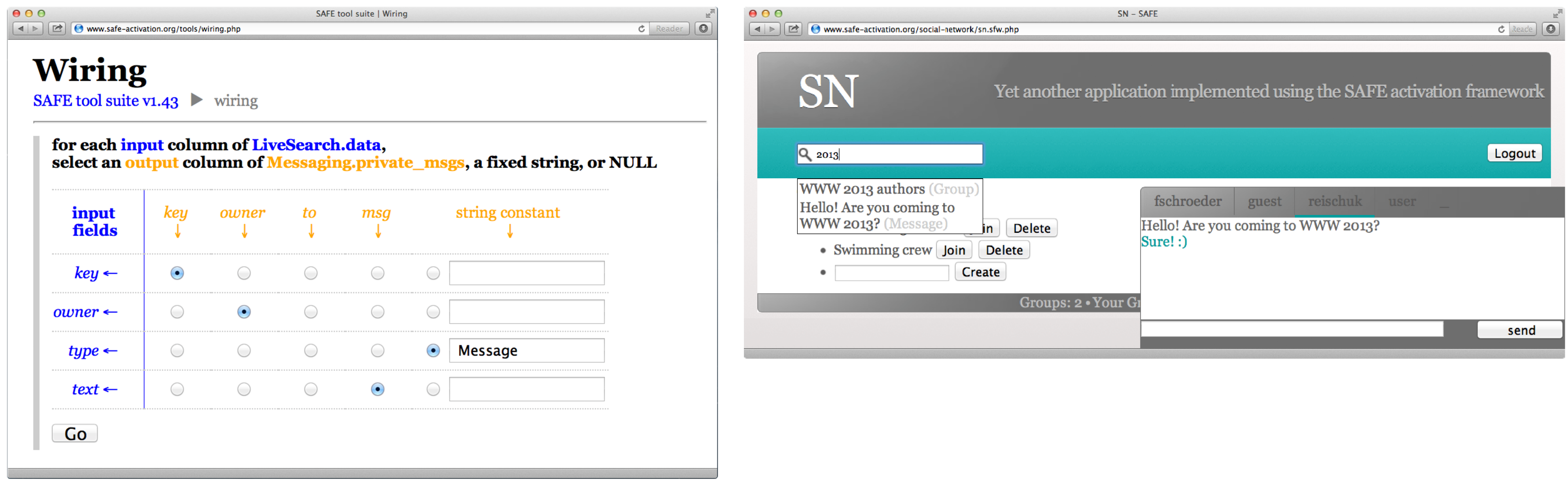}{0px 0px 0px 0px}{1}{Screenshots: Schema matching
between \emph{Messaging} and \emph{LiveSearch} (left), and the
extended application in which the external \emph{LiveSearch} f-unit
displays results obtained from customizable sources (right).}

\myparagraph{Adding Functionality}
Given the initial application, we will now add a common
\emph{incremental search} functionality. By this means, the f-unit
\emph{LiveSearch} monitors a text input field for typing events,
searches all its available datasets for the input pattern, and
displays matching rows. As introduced in \autoref{implementation}, we
have equipped \emph{LiveSearch} with an input table \texttt{data}
that can be wired to output tables of other f-units. 
\begin{lsthere}
    INPUT TABLE <latex>\textcolor{darkblue}{data}\!</latex> ( text  TEXT
                      type  VARCHAR(20)
                      key   KEY
                      owner OWNER )
\end{lsthere}
The input table
has two main data fields: \texttt{text} for arbitrary textual content
(e.g., chat messages, group titles, poll descriptions), and
\texttt{type} for an informal description of the search source type
(e.g., messages, groups, polls).

By virtue of this input table, \emph{LiveSearch} is able to search
arbitrary data sets, even for data sources that are provided by
f-units that were not known before, or by f-units that might come up
in the future. At runtime, \emph{LiveSearch} compares these data
sources with the search patterns entered in \emph{LiveSearch}'s
search input field:
\begin{lsthere}
    <input type="text" name="<latex>\textcolor{darkgreen}{search}</latex>" id="searchField">
\end{lsthere}
\emph{LiveSearch} issues queries against its input table
\texttt{data} for every \texttt{keyup}-event of the search field and
activates corresponding instances of the f-unit
\emph{LiveSearchResults}:\footnote{The activation tag in the code
snippet is part of \SAFE's modeling language SFW, an extended
HTML-based declarative programming language that allows for concise
incorporation of HTML constructs, JavaScript events, and SQL queries.}

\begin{lsthere}
    <activate:LiveSearchResults
     query="SELECT text AS result, type AS info
            FROM <latex>\textcolor{darkblue}{data}</latex>
            WHERE '<latex>\textcolor{darkgreen}{\$\#search}</latex>'<>'' AND
              LOWER(text) LIKE LOWER(CONCAT('%',REPLACE('<latex>\textcolor{darkgreen}{\$\#search}</latex>',' ','%'),'%'))"
     refresh="searchField.keyup" />
\end{lsthere}


\myparagraph{Wiring}
In the social network setting, the search engine shall include the
groups of the social network in its search results. In order to
provide \emph{LiveSearch} with the actual group names, the wiring of
\emph{Groups}\texttt{.all\_groups} into
\emph{LiveSearch}\texttt{.data} maps \texttt{key} $\mapsto$
\texttt{key}, \texttt{name} $\mapsto$ \texttt{text}, the constant
\texttt{'Group'} $\mapsto$ \texttt{type}, and \texttt{owner}
$\mapsto$ \texttt{owner}. Furthermore, upon integration of
\emph{Messaging}, the new feature of searching in both sent and
received messages can be stated by the wiring shown in
\autoref{screenshots} (left-hand side): \texttt{key} $\mapsto$
\texttt{key}, \texttt{msg} $\mapsto$ \texttt{text}, the constant
\texttt{'Message'} $\mapsto$ \texttt{type}, and \texttt{owner}
$\mapsto$ \texttt{owner}.


\myparagraph{Evaluation}
The right-hand side of \autoref{screenshots} shows the resulting application: a
wired input table allows \emph{LiveSearch} to display search results
generically for datasets of both \emph{Groups} and \emph{Messaging}.
The wiring of \emph{Groups}\texttt{\!.\!all\_groups} and
\emph{Messaging}\texttt{\!.\!private\_msgs} into
\emph{LiveSearch}\texttt{.data} results in a safe setting that
reflects the modularity and extensibility par\-a\-digms, as depicted
above.
The implementation of \emph{LiveSearch} benefits from various
features and concepts that are offered by the described extensions of
\SAFE.
For instance, the result set of \emph{LiveSearch} can be arbitrarily
augmented at ``run-time'', and the wiring allows for easy integration
of new functionality into an existing app ecosystem, without
affecting already established apps. Collaboration across f-units thus
only relies on a sufficiently generic interface of all involved
f-units, formed by input and output tables.
Furthermore, \emph{LiveSearchResults} --- or any other involved
f-unit --- can be replaced by means of extensibility with respect to
both presentation and functionality, allowing for augmenting the
application in unforeseen directions.
In addition, even though \emph{Messaging} publishes
privacy-sen\-si\-tive data, \emph{Messaging} is able to bind datasets
to appropriate invariants and thus has full control over which data
might possibly be presented to other users. Consequently, the impact
of an extended malicious f-unit (for instance \emph{LiveSearch}) on
the overall system security is limited to the abuse of the malicious
f-unit's very own or received datasets.
Finally, \SAFE's activation tag \texttt{<activate..>} with the
attributes \texttt{query} and \texttt{refresh} allows for a
straight-forward implementation without the need for cumbersome
user-defined AJAX handling --- the resulting gain is a high
functionality/LoC ratio with all its implied desirable correctness
and security properties.

  \section{Additional Related Work}
\label{sec:related}

%
%
Similar to \SAFE, the \textit{WebRatio} development environment
\cite{Acerbis2004:WebRatio,Brambilla2007:WebML-WebRatio}, and, in particular,
the \textit{WebML} language \cite{WebML2000a}, follow the approach of building web
applications by composing and connecting so-called \emph{content units}. These
units are modeled and structured by means of an abstract, data-centric
description and thus strongly resemble data dependencies as well as the actual
data representation layout. Our focus, however, is to address the possibility
of incorporating third-party code, which requires the presence of appropriate
security mechanisms.
%
%
\textit{SMash}~\cite{DeKeukelaere2008:SMash} addresses the task of
combining data and code from different origins in standard
(i.e., unmodified) browsers based on HTML \texttt{iframes}.
The system establishes secured communication channels
between the different components. However, there is no
unified database support for the different components, and
there is no enforced principal model as required in the case
of data-driven applications with sensitive user data,
ownership and provenance. Instead of presenting an
iframe-based implementation, our model provides a novel
multi-dimensional privilege model suitable for extensible
mashup applications.
Moreover, in contrast to \cite{DeKeukelaere2008:SMash} and
also
\cite{Magazinius2010:LatticeBasedApproach,BJM09:frames}, our
model does not assume different origins per each integrated
component. Technically, the origin is defined by the triple
scheme/servername/port. Instead, untrusted code is
integrated on a single server that delivers all necessary
code at once.
%
As in our model, the \textit{ServiceOS} operating system
\cite{Wang2009:Convergence} considers web application
components (as well as desktop applications) as first-class
principals. Multi-application sharing is brought to web
applications with cross-principal protection and resource
management. The approach is different in that our approach
directly augments the developed applications with
functionality and security, instead of providing an
operating system. In particular, applications in our model run in
every standard browser and thus do not require a certain
environment to be installed at the client side.

  \section{Conclusions}\label{sec:conclusions}

We have proposed a novel extensibility mechanism which is
designed for the implementation of extensibility for
existing web applications. Possibly untrusted
components can be integrated in an app ecosystem in a secure
and privacy-friendly manner. Our multi-dimensional principal
model provides a clean component abstraction, thereby
impeding undesired component access and ensuring that no
undesired information flow takes place between application
components. We have instantiated our model in the \SAFE
activation framework, resulting in a novel methodology that
is specifically designed for the newly emerging needs of
extensibility in application ecosystems. We have illustrated
the convenient usage of our techniques by showing how to
securely extend an existing social network application.

  \bibliographystyle{abbrv}
  \bibliography{src/bib,src/links}
\end{document}